%% file: paper_resub_Feb13.tex
\documentclass[twocolumn,showpacs,aps,prd,fleqn]{revtex4}
\usepackage{fleqn}
\usepackage{graphicx}
\usepackage{dcolumn}
\usepackage{amsmath}
\usepackage{epsfig}
\usepackage{relsize}
\usepackage{multirow}
\RequirePackage{xspace}
\input babarsym

\def\ie{{\it i.e.}}

\def\Dkkpi    {\ensuremath{D^{+}\to K^{+}K^{-}\pi^{+}}\xspace}
\def\kkpi     {\ensuremath{K^{+}K^{-}\pi^{+}}\xspace}
\def\kkpg     {\ensuremath{K^{+}K^{-}\pi^{+}\gamma}\xspace}
\def\kk       {\ensuremath{K^{+}K^{-}}\xspace}
\def\kpi      {\ensuremath{K^{-}\pi^{+}}\xspace}
\def\mkk      {\ensuremath{m(K^{+}K^{-})}\xspace}
\def\mkpi     {\ensuremath{m(K^{-}\pi^{+})}\xspace}

\def\coscmnox {\ensuremath{\cos(\theta_{\mathrm{CM}})}}
\def\coscm    {\ensuremath{\cos(\theta_{\mathrm{CM}})}\xspace}
\def\pcm      {\ensuremath{p_{\mathrm{CM}}}\xspace}
\def\msqkk    {\ensuremath{m^{2}(K^{+}K^{-})}\xspace}
\def\msqkpi   {\ensuremath{m^{2}(K^{-}\pi^{+})}\xspace}
\def\CP       {\ensuremath{C\!P}\xspace}

\def\CPV      {\ensuremath{C\!PV}\xspace}

\def\ACP      {\ensuremath{A_{C\!P}}\xspace}
\def\etal     {{\it et al.}}
\def\Kres     {\ensuremath{\bar{K}^*(892)^0}}
\def\phires   {\ensuremath{\phi(1020)}}
\def\Dstarplus {\ensuremath{D^{*+}}}

\newcommand{\gevcccc}{\ensuremath{{\mathrm{\,Ge\kern -0.1em V^2\!/}c^4}}\xspace}
\newcommand{\BABARPubYear}    {12}
\newcommand{\BABARPubNumber}  {014}

\newcommand{\SLACPubNumber} {15077}

\def\figurebox#1#2#3{%
    \def\arg{#3}%
    \ifx\arg\empty
    {\hfill\vbox{\hsize#2\hrule\hbox to #2{\vrule\hfill\vbox to #1{\hsize#2\vfill}\vrule}\hrule}\hfill}%
    \else
    {\hfill\epsfbox{#3}\hfill}%
    \fi}

\begin{document}
\preprint{\babar-PUB-\BABARPubYear/\BABARPubNumber} 
\preprint{SLAC-PUB-\SLACPubNumber} 

\begin{flushleft}
\babar-PUB-\BABARPubYear/\BABARPubNumber\\
SLAC-PUB-\SLACPubNumber\\
\end{flushleft}

\title{
{\large \bf
Search for direct \CP violation in singly Cabibbo-suppressed {\boldmath$D^{\pm}\rightarrow K^{+}K^{-}\pi^{\pm}$} decays} 
}

\input authors_jul2012
\setcounter{footnote}{0}

\begin{abstract}
We report on a search for direct \CP violation in the singly
Cabibbo-suppressed decay \Dkkpi using a data sample of 476 \invfb\ of
$e^+e^-$ annihilation data 
accumulated with the \babar\ detector at the SLAC PEP-II
electron-positron collider running at and just below the energy of the
\Y4S resonance. The integrated \CP-violating decay rate asymmetry \ACP is
determined to be (0.37 $\pm$ 0.30 $\pm$ 0.15)$\%$. Model-independent and
model-dependent Dalitz plot analysis techniques are used to search for
\CP-violating asymmetries in the various intermediate states. We find no
evidence for \CP-violation asymmetry.
\end{abstract}

\maketitle
\section{Introduction}
\label{sec:intro}
Searches for \CP violation ({\CPV}) in charm meson decays provide a
probe of physics beyond the Standard Model (SM).  Singly
Cabibbo-suppressed (SCS) decays can exhibit direct \CP asymmetries due
to interference between tree-level transitions and $|\Delta C| = 1$
penguin-level transitions if there is both a strong and a weak phase
difference between the two amplitudes. In the SM, the resulting
asymmetries are suppressed by
$\mathcal{O}(|V_{cb}V_{ub}/V_{cs}V_{us}|)\sim10^{-3}$, where
$V_{\mathrm{ij}}$ are elements of the Cabibbo-Kobayashi-Maskawa
quark-mixing matrix~\cite{CKM}. A larger measured value of the \CP
asymmetry could be a consequence of the enhancement of penguin
amplitudes in $D$ meson decays due to final-state
interactions~\cite{GronauRosner,ChengChiang}, or of new
physics~\cite{GKN07,AMP08}.

The LHCb and CDF Collaborations recently reported evidence for a
non-zero \CP\ asymmetry in the difference of the time-integrated $D^0\to
\pi^+\pi^-$ and $D^0\to K^+K^-$ decay rates~\cite{LHCb, CDF}. Searches
for {\CPV} in other SCS decays with identical transitions $c\to
ud\bar{d}$ and $c\to u s \bar{s}$ are relevant to an understanding of
the origin of {\CPV}~\cite{GKZ12, GIP12, FMS12}.

We present here a study of the SCS decay \Dkkpi~\cite{cc}, which is
dominated by quasi-two-body decays with resonant intermediate
states. This allows us to probe the Dalitz-plot substructure for
asymmetries in both the magnitudes and phases of the intermediate
states. The results of this study include a measurement of the
integrated \CP asymmetry, the \CP asymmetry in four regions of the
Dalitz plot, a comparison of the binned $D^+$ and $D^-$ Dalitz plots, a
comparison of the Legendre polynomial moment distributions for the \kk
and \kpi systems, and a comparison of parameterized fits to the Dalitz
plots.  Previous measurements by the \mbox{CLEO-c} Collaboration found no
evidence for {\CPV} in specific two-body amplitudes or for the integrals
over the entire phase-space~\cite{CLEO-eff}.  The LHCb Collaboration
also finds no evidence for {\CPV} in a model-independent search
\cite{LHCb2}.

\section{The \babar\ Detector and Data Sample}
\label{sec:det}
The analysis is based on a sample of electron-positron annihilation data
collected at and just below the energy of the \Y4S resonance with the
\babar\ detector at the SLAC \pep2\
collider, corresponding to an integrated luminosity of 476\invfb.  The
\babar\ detector is described in detail elsewhere~\cite{Aubert:2001tu}.
The following is a brief summary of the detector subsystems important to
this analysis. Charged-particle tracks are detected, and their momenta
measured, by means of the combination of a 40-layer cylindrical drift
chamber (DCH) and a five-layer silicon vertex tracker (SVT), both
operating within a 1.5-T solenoidal magnetic field.  Information from a
ring-imaging Cherenkov detector (DIRC) and specific energy-loss
measurements ($dE/dx$) in the SVT and DCH are used to identify charged
kaon and pion candidates.

For various purposes described below we use samples of Monte Carlo (MC)
simulated events generated using the JETSET~\cite{jetset} program. These
events are passed through a detector simulation based on the Geant4
toolkit~\cite{Agostinelli:2002hh}. Signal MC events refer to $\Dkkpi$
decays generated using JETSET as well as $\Dkkpi\gamma$ decays generated
using JETSET in combination with the PHOTOS~\cite{photos} program. In
all cases when we simulate particle decays we include EvtGen.~\cite{evtgen}

\section{Event Selection and {\boldmath$\Dkkpi$} Reconstruction}
\label{sec:evtsel}
The three-body $\Dkkpi$ decay is reconstructed from events having at
least three tracks with net charge $+1$. 
Two oppositely charged tracks must be consistent with the kaon
hypothesis. Other charged tracks are assumed to be pions. To improve
particle identification performance, there must be at least one photon
in the DIRC associated with each track. Contamination from electrons is
significantly reduced by means of $dE/dx$ information from the DCH.
Pion candidates must have transverse momentum $p_T > 300$ \mevc. For
lower $p_T$ values, tracks are poorly reconstructed. Also, for lower
$p_T$, differences in the nuclear cross sections for positively charged
and negatively charged particles can lead to asymmetries.  We form the
invariant mass of \kkpi\ candidates and require it to lie within
1.82-1.92\gevcc. The three tracks must originate from a common vertex,
and the vertex-constrained fit probability ($P_{\mathrm{vtx}}$) must be
greater than 0.5\%. The momentum in the CM frame (\pcm) of the resulting
$D$ candidate must lie within the interval [2.4, 5.0] \gevc. The lower
limit on \pcm reduces background from $B$ decays by preferentially
selecting $e^+e^-\rightarrow c\bar{c}$ events; this has traditionally
been the way to reduce combinatoric background due to $B$ decays.
To remove background
from misidentified $\Dstarplus\rightarrow\Dz\pi^{+}$ decays, we require
$m(\kkpi) - m(\kpi) - m(\pi^{+}) > 15 \mevcc$, where the pion and kaon
masses are set to the nominal values~\cite{PDG10}.  Finally, for events
with multiple $D^\pm$ candidates, the combination with the largest value
of $P_{\mathrm{vtx}}$ is selected.  We perform a separate kinematic fit
in which the \Dpm mass is constrained to its nominal value~\cite{PDG10}.
The result of the fit is used in the Dalitz plot and moments analyses
described below.

To aid in the discrimination between signal and background events, we use
the joint probability density function (PDF) for $L_{\mathrm{xy}}$, the
distance between the primary event vertex and the $D$ meson decay vertex
in the plane transverse to the beam direction, and \pcm, to form a
likelihood ratio:
\begin{equation}
\mathrm{R}_{\cal{L}} = 
  \frac{P_{\mathrm{s}}(p_{\mathrm{CM}})P_{\mathrm{s}}(L_{\mathrm{xy}})}{P_{\mathrm{s}}(p_{\mathrm{CM}})P_{\mathrm{s}}(L_{\mathrm{xy}})
  + P_{\mathrm{b}}(p_{\mathrm{CM}})P_{\mathrm{b}}(L_{\mathrm{xy}})}. 
\end{equation}
Since the two variables have little correlation we construct the
two-dimensional PDF as simply the product of their one-dimensional PDFs;
these one-dimensional PDFs for signal ($P_{\mathrm{s}}$) and background
($P_{\mathrm{b}}$) are estimated from data. The background PDFs are
determined from events in the \Dp~mass sidebands, while those for the
signal are estimated from events in the \Dp~signal region after
background is subtracted using estimates from the sidebands. The signal
region is defined by the $m(\kkpi)$ interval 1.86-1.88 \gevcc, while the
sideband regions are the 1.83-1.84 \gevcc and 1.90-1.91 \gevcc
intervals.
The selection on $\mathrm{R}_{\cal{L}}$ is adjusted to maximize signal
significance, 
and the resulting signal is fairly pure (see Fig.~\ref{fig:massfit} in
Sec.~\ref{sec:massfit}).

The reconstruction efficiency for $\Dp$ decays is determined from a
sample of MC events in which the decay is generated according to phase
space (\ie, the Dalitz plot is uniformly populated). To parameterize
the selection efficiency, we use the distribution of reconstructed
events as a function of the cosine of the polar angle of the $D$ meson
in the CM frame [\coscmnox] and the \msqkpi versus \msqkk Dalitz
plot. The selection efficiency is determined as the ratio of
$N_{\mathrm{Reco}}/N_{\mathrm{Gen}}$ in intervals of \coscm and
separately in intervals of the Dalitz plot, where $N_{\mathrm{Reco}}$ is
the number of selected events in an interval and $N_{\mathrm{Gen}}$ is
the number of events generated in the same interval. The binned
Dalitz-plot efficiency is parameterized with a feed-forward artificial
neural network (ANN)~\cite{mlpfit} consisting of two hidden layers with
three and five nodes. Use of an ANN procedure allows us to adequately
model the efficiency near the edges of the Dalitz plots. The ANN
efficiency function is tested by creating separate training and
validation samples, which are satisfactorily fit by the ANN.

\section{Corrections to Simulated Events}
\label{sec:corrmc}
In order to describe accurately the reconstruction efficiency, we apply
corrections to the reconstructed MC events to account for known
differences between simulated events and data. The differences arise in
the reconstruction asymmetry of charged-pion tracks, and in the
production model for charm mesons. Differences in kaon particle
identification efficiency have a negligible asymmetry effect since the
$K^+$ and $K^-$ are common to $D^+$ and $D^-$ decays.

To correct the production model used in the simulation, we construct the
ratio of the two-dimensional \pcm versus \coscm PDFs between 
data and simulation and apply this ratio as a correction to the
reconstructed MC events before calculating the efficiency.  For this
procedure the signal PDF for data is background-subtracted while the
signal MC events are weighted by the Dalitz plot amplitude-squared
determined from data (see Sec.~\ref{sec:dpamp}).

\begin{figure}[!tb]
\begin{center}
\includegraphics[width=0.45\textwidth]{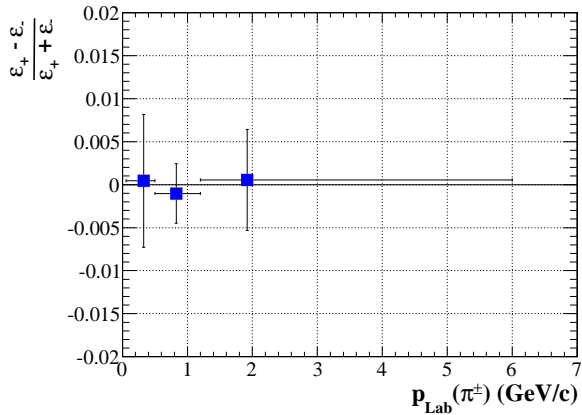}
\vspace{-0.3cm}
\caption{(color online) Charged pion tracking efficiency asymmetry
  (defined in Eq.~(\ref{eq:trkasym})) as a function of the pion momentum
  in the laboratory frame determined from the decays of $\tau$
  leptons. The horizontal error bars indicate the range of pion
  momentum~\cite{BaBarTrk}.}
\label{fig:trkasym}
\vspace{-0.7cm}
\end{center}
\end{figure} 

To correct for differences in the reconstruction asymmetry of
charged-pion tracks, we use a sample of $\epem\to\tau^+\tau^-$ events in
which one $\tau$ decays leptonically via
$\tau^\pm\rightarrow\mu^\pm\nu_\mu\nu_\tau$ while the other $\tau$
decays hadronically via $\tau^\mp\rightarrow h^\mp h^\mp
h^\pm\nu_\tau$. We tag events with a single isolated muon on one side of
the event and reconstruct the hadronic $\tau$ decay in the opposite
hemisphere. We refer to this sample as the ``Tau31'' sample. We further
require two of the three hadrons to have an invariant mass consistent
with the rho mass to within 100 \mevcc. Due to tracking inefficiencies,
tau decays to three tracks are sometimes reconstructed with only two
tracks. We use the two-dimensional distributions of
$\cos\theta_{\pi\pi}$ and $p_{T_{\pi\pi}}$ (with respect to the beam
axis) of the rho-decay pions for 2-hadron and 3-hadron events to
determine the pion inefficiency and asymmetry. We allow for a different
efficiency for positive and negative tracks ($\varepsilon_{\pm}$) by
introducing the asymmetry $a(p_{\mathrm{Lab}})$ as a function of pion
laboratory momentum ($p_{\mathrm{Lab}}$):

\begin{equation}\label{eq:trkasym}
    a(p_{\mathrm{Lab}}) = \frac{\varepsilon_+(p_{\mathrm{Lab}}) - 
	\varepsilon_-(p_{\mathrm{Lab}})}{\varepsilon_+(p_{\mathrm{Lab}}) + \varepsilon_-(p_{\mathrm{Lab}})}.
\end{equation}

The results for $a(p_{\mathrm{Lab}})$ are shown in
Fig.~\ref{fig:trkasym}: the average value for $0 < p_{\mathrm{Lab}} < 4$
\gevc\ is (0.10 $\pm$ 0.26)$\%$, which is consistent with
zero~\cite{BaBarTrk}. We use linear interpolation between data points,
or extrapolation beyond the first and last data points, to obtain the
ratio of track-efficiency asymmetries between data and MC as a function
of momentum. This ratio is then used to correct track efficiencies
determined from signal MC.

\section{\boldmath Integrated $CP$ Asymmetry as a Function of \coscm}
\label{sec:intgacp}
\begin{figure}[!tb]
\begin{center}
\includegraphics[width=0.45\textwidth]{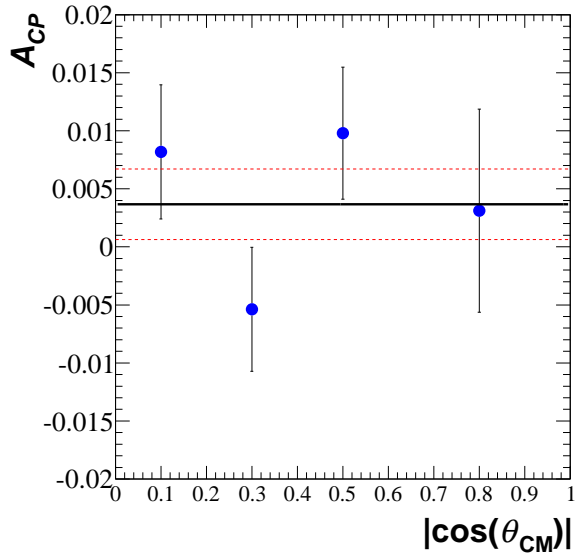}
\vspace{-0.3cm}
\caption{(color online) \CP asymmetry as a function of $|\coscm|$.  The
  solid line represents the central value of \ACP and the dashed lines
  the $\pm1$ standard deviation statistical uncertainty, determined from
  a $\chi^2$ fit to a constant value.}
\label{fig:acp}
\vspace{-0.7cm}
\end{center}
\end{figure}

The production of \Dp (and \Dm) mesons from the $\epem\to\ccbar$ process
is not symmetric in \coscm; this forward-backward (FB) asymmetry,
coupled with the asymmetric acceptance of the detector, results in
different yields for \Dp and \Dm events. The FB asymmetry, to first
order, arises from the interference of the separate annihilation
processes involving a virtual photon and a $Z^{0}$ boson. We define the
charge asymmetry $A$ in a given interval of \coscm by
\begin{equation}
  A(\coscm) \equiv \frac{ N_{D^{+}}/\epsilon_{D^{+}}  - N_{D^{-}}/\epsilon_{D^{-}}}
    {N_{D^{+}}/\epsilon_{D^{+}} + N_{D^{-}}/\epsilon_{D^{-}}},
\end{equation}
where $N_{D^\pm}$ and $\epsilon_{D^\pm}$ are the yield and efficiency,
respectively, in the given \coscm bin. 
We remove the FB asymmetry by averaging $A$ over four intervals
symmetric in \coscm, \ie, by evaluating
\begin{equation}
  \label{eqn:acp}
  A_{CP} \equiv \frac{A(\coscm) + A(-\coscm)}{2}.
\end{equation}
The interval boundaries in \coscm are defined as [0, 0.2, 0.4, 0.6,
  1.0]. The \Dpm yields are determined from fits to the reconstructed
$K^{\pm}K^{\mp}\pi^{\pm}$ mass distributions, as described in
Sec.~\ref{sec:massfit}.  This technique has been used in previous
\babar~measurements in both three-body and two-body decays
~\cite{Aubert:2008kkpipi,Aubert:2008hhpi0,delAmoSanchez:2011kspi}.  The
weighted average of values obtained using Eq.~(\ref{eqn:acp}) is $\ACP = $
$(0.37\pm0.30\pm0.15)\%$, where the uncertainties are statistical and
systematic, respectively, with a probability of 21\%\ that the
asymmetries are null in all four intervals (Fig.~\ref{fig:acp}).

\section{\boldmath$\Dp$ Mass Fit}
\label{sec:massfit}

\begin{figure}[!tb]
\begin{center}
\includegraphics[width=0.5\textwidth]{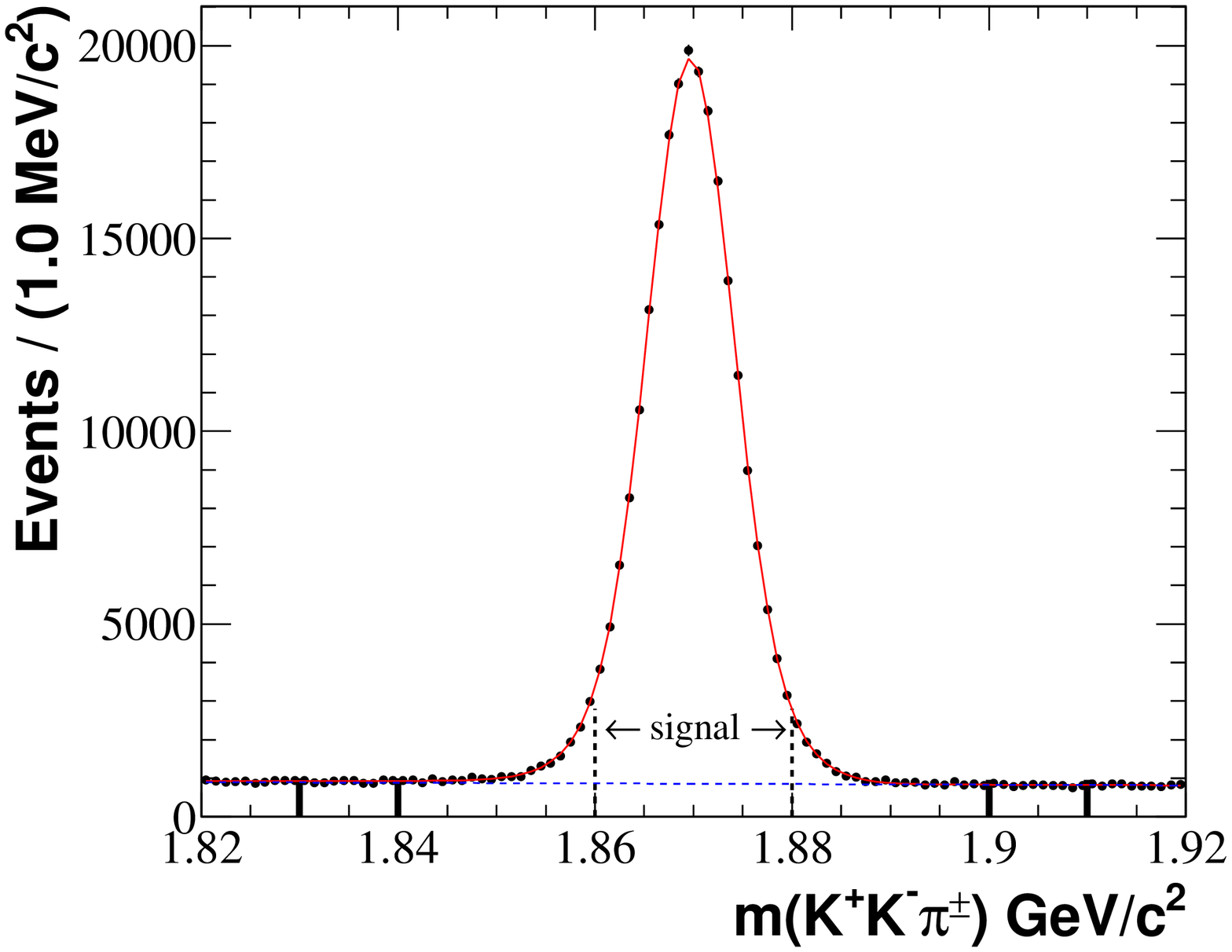}
\includegraphics[width=0.5\textwidth]{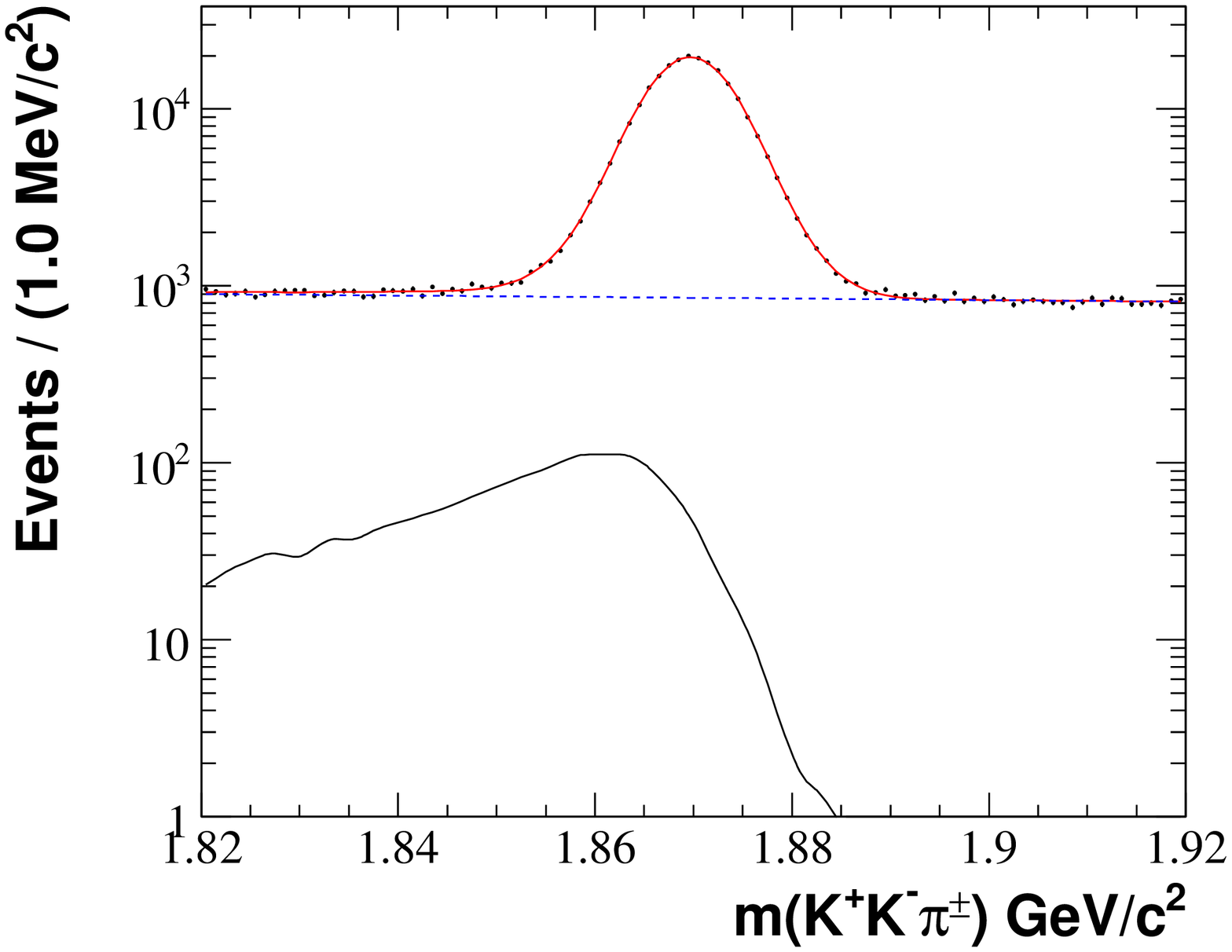}
\vspace{-0.3cm}
\caption{(color online) Combined reconstructed invariant mass
  distribution $m(\kk\pi^{\pm})$ and projection of the fit result. The
  points show the data, the solid curve the fit model, and the dashed
  curve shows the background PDF. The signal region is indicated by the
  dashed vertical lines, and the sideband regions by the solid vertical
  lines. The lower figure shows the fit on a logarithmic scale with the
  radiative component of the signal PDF shown separately as a smooth
  curve.}
\label{fig:massfit}
\vspace{-0.3cm}
\end{center}
\end{figure} 

The \kkpi mass distribution is fitted with a double-Gaussian function
with a common mean and a linear background (Fig.~\ref{fig:massfit}),
plus a function describing radiative decays $\Dkkpi\gamma$. The PDF for
radiative decays is obtained from the reconstructed mass distribution of
$\kkpg$ events selected at the generator-level in our MC additionally
convolved with a Gaussian of width 2.26 MeV/$c^2$, and accounts for
1.5\%\ of the signal. The fit to data gives a \Dp mass value of 1869.70
$\pm$ 0.01 \mevcc, where the uncertainty is statistical only.  The
signal region is defined to lie within $\pm2\sigma_{\Dp}$ of the peak,
where $\sigma_{\Dp} = \sqrt{f_1\sigma_1^2 + (1-f_1)\sigma_2^2}$ is
5.04~\mevcc, and contains a total of 227874 events; $\sigma_1(\sigma_2)$
is the standard deviation of the first (second) Gaussian component and
$f_1 = 0.63$ is the fraction of the signal in the first Gaussian
component. Separate fits to the $\kk\pi^+$ and $\kk\pi^-$ distributions
yield $N_{\Dp} = 113037 \pm 469$ and $N_{\Dm} = 110663 \pm 467$ events,
respectively. The ratio of efficiency-corrected yields ($N/\epsilon$) is
$R \equiv \frac{N_{\Dp}/\epsilon_{\Dp}}{N_{\Dm}/\epsilon_{\Dm}} = 1.020
\pm 0.006$. This ratio is used to account for remaining asymmetries that
arise from physics- or detector-related processes, such as an
insufficiently accurate simulation of the FB asymmetry or a residual
detector asymmetry. Also, it is a less accurate measure of the asymmetry
when the efficiency varies significantly as a function of
$\coscm$, as for our experiment.

\section{\boldmath Model-independent searches for $CP$ violation in the Dalitz plots}
\label{sec:moments}
\begin{figure*}[!tb]
\begin{center}
\includegraphics[width=0.98\textwidth]{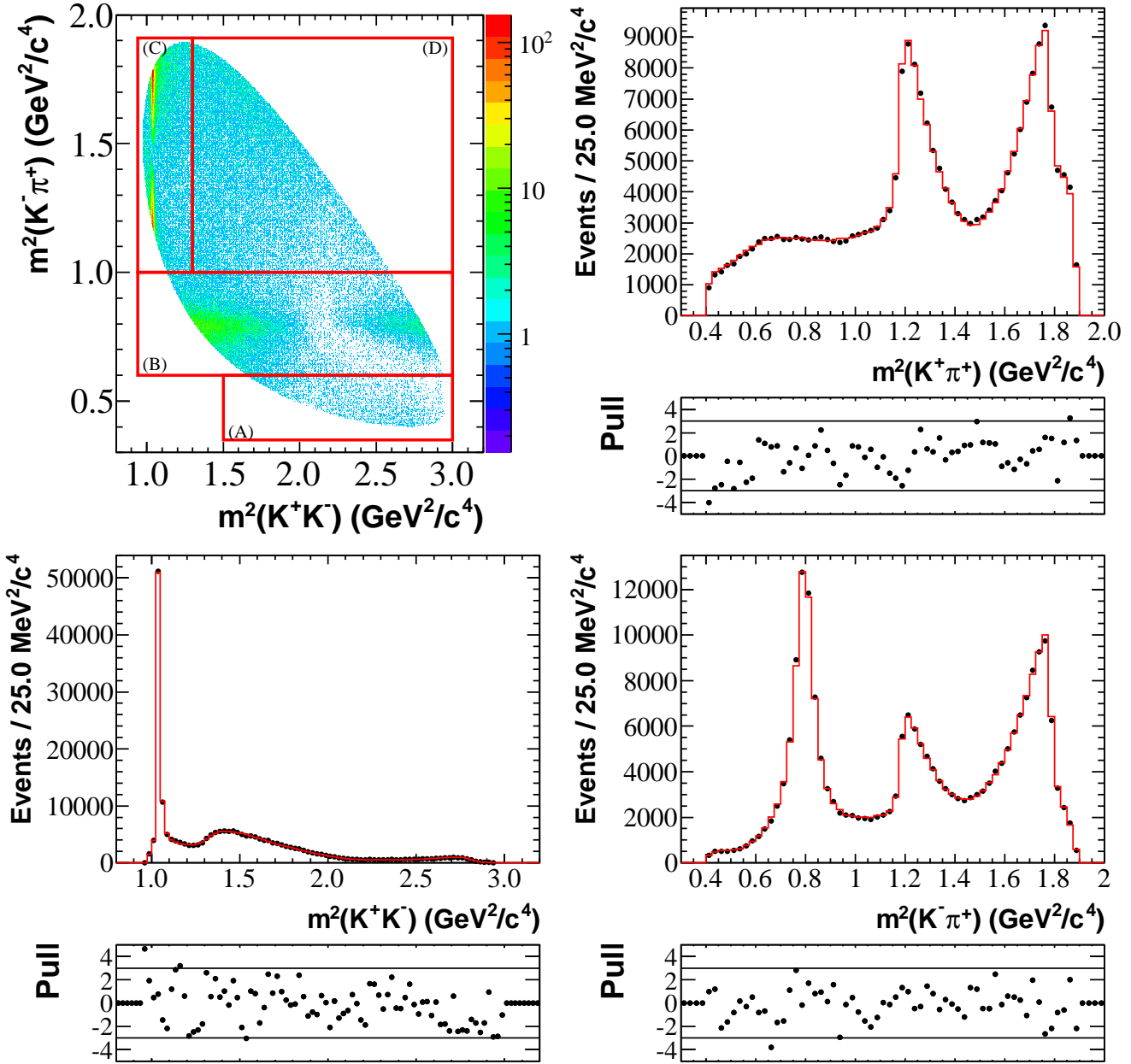}
\vspace{-0.3cm}
\caption{(color online) $\Dpm\rightarrow \kk\pi^{\pm}$ Dalitz plot and
  fit projections assuming no {\CPV}, with the regions used for
  model-independent comparisons indicated as boxes. The A/B boundary is
  at $m_{K\pi} = 0.6 \; $GeV$^2/c^4$, the B/C boundary at $m_{K\pi} =
  1.0 \; $GeV$^2/c^4$, and the C/D boundary at $m_{KK} = 1.3 \;
  $GeV$^2/c^4$. In the fit projections, the data are represented by
  points with error bars and the fit results by the histograms. The
  normalized residuals below each projection, defined as
  $(N_{\mathrm{Data}} - N_{\mathrm{MC}})/\sqrt{N_{\mathrm{MC}}}$, lie
  between $\pm5\sigma$.  The horizontal lines correspond to $\pm
  3\sigma$.}
\label{fig:dpfit}
\vspace{-0.7cm}
\end{center}
\end{figure*}

Model-independent techniques to search for \CP violation in the Dalitz
plots are presented in Ref.~\cite{Aubert:2008hhpi0}. The techniques
include a comparison of the moment distributions, and the asymmetry in
the \Dp and \Dm yields in various regions of the Dalitz plot.  We scale
the \Dm\ yields by the factor $R$ described in Sec.~\ref{sec:massfit}.
By applying this correction, we remove residual detector-induced
asymmetries and decouple, as far as possible, the search for {\CPV} in
the Dalitz plot from the search for {\CPV} integrated over the phase
space, which was described in Sec.~\ref{sec:intgacp}. We measure the \CP
asymmetry in the four regions of the Dalitz plot labeled A, B, C, and D
in Fig.~\ref{fig:dpfit}.  We report the fitted yields, average Dalitz
plot efficiencies, and \CP asymmetries in
Table~\ref{tab::binnedDPAcp_ub}.

\begin{table*}
\begin{center}
\caption{Yields, efficiencies, and \CP asymmetry in the regions of the
Dalitz plot shown in Fig.~\ref{fig:dpfit}. For the \CP asymmetry, the
first uncertainty is statistical and the second is systematic.}
\begin{tabular}{lccccc}
\hline 
\hline
Dalitz plot region & $N(\Dp)$ & $\epsilon(\Dp)[\%]$ & $N(\Dm)$ & $\epsilon(\Dm)[\%]$ & $\ACP[\%]$ \\ 
\hline
(A) Below $\bar{K}^{*}(892)^0$ & 1882 $\pm$ 70 & 7.00 & 1859 $\pm$ 90 & 6.97 & $-0.7 \pm 1.6 \pm 1.7$\\ 
(B) $\bar{K}^{*}(892)^0$ & 36770 $\pm$ 251 & 7.53 & 36262 $\pm$ 257 & 7.53 & $-0.3 \pm 0.4 \pm 0.2$\\ 
(C) $\phi(1020)$ & 48856 $\pm$ 289 & 8.57 & 48009 $\pm$ 289 & 8.54 & $-0.3 \pm 0.3 \pm 0.5$\\ 
(D) Above $\bar{K}^{*}(892)^0$ and $\phi(1020)$ & 25616 $\pm$ 244 & 8.01 & 24560 $\pm$ 242 & 8.00 & $1.1 \pm 0.5 \pm 0.3$\\
\hline
\hline
\end{tabular} 
\label{tab::binnedDPAcp_ub}
\end{center}
\end{table*}

\begin{figure}[!tb]
\begin{center}
\includegraphics[width=0.45\textwidth]{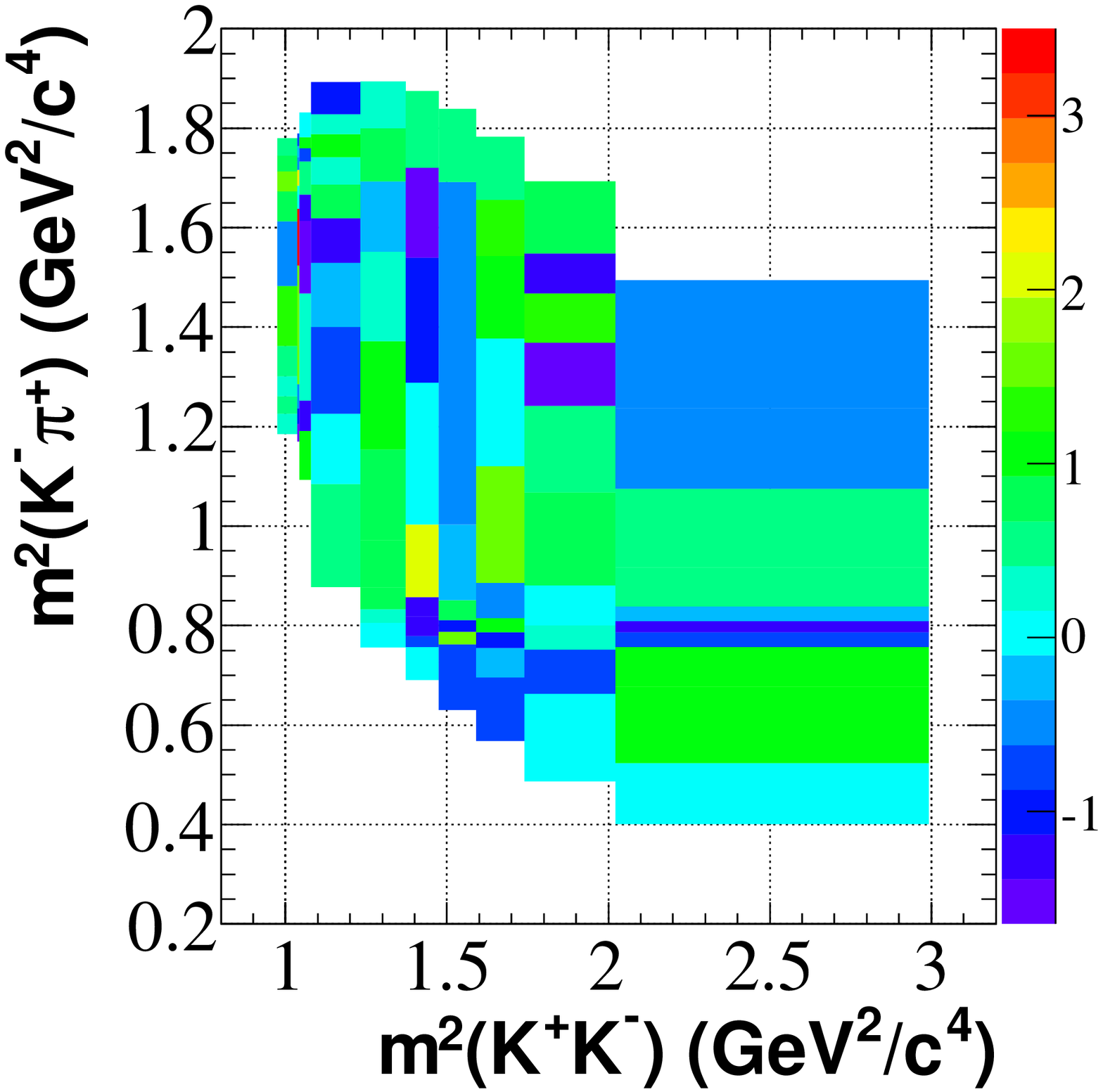}
\includegraphics[width=0.45\textwidth]{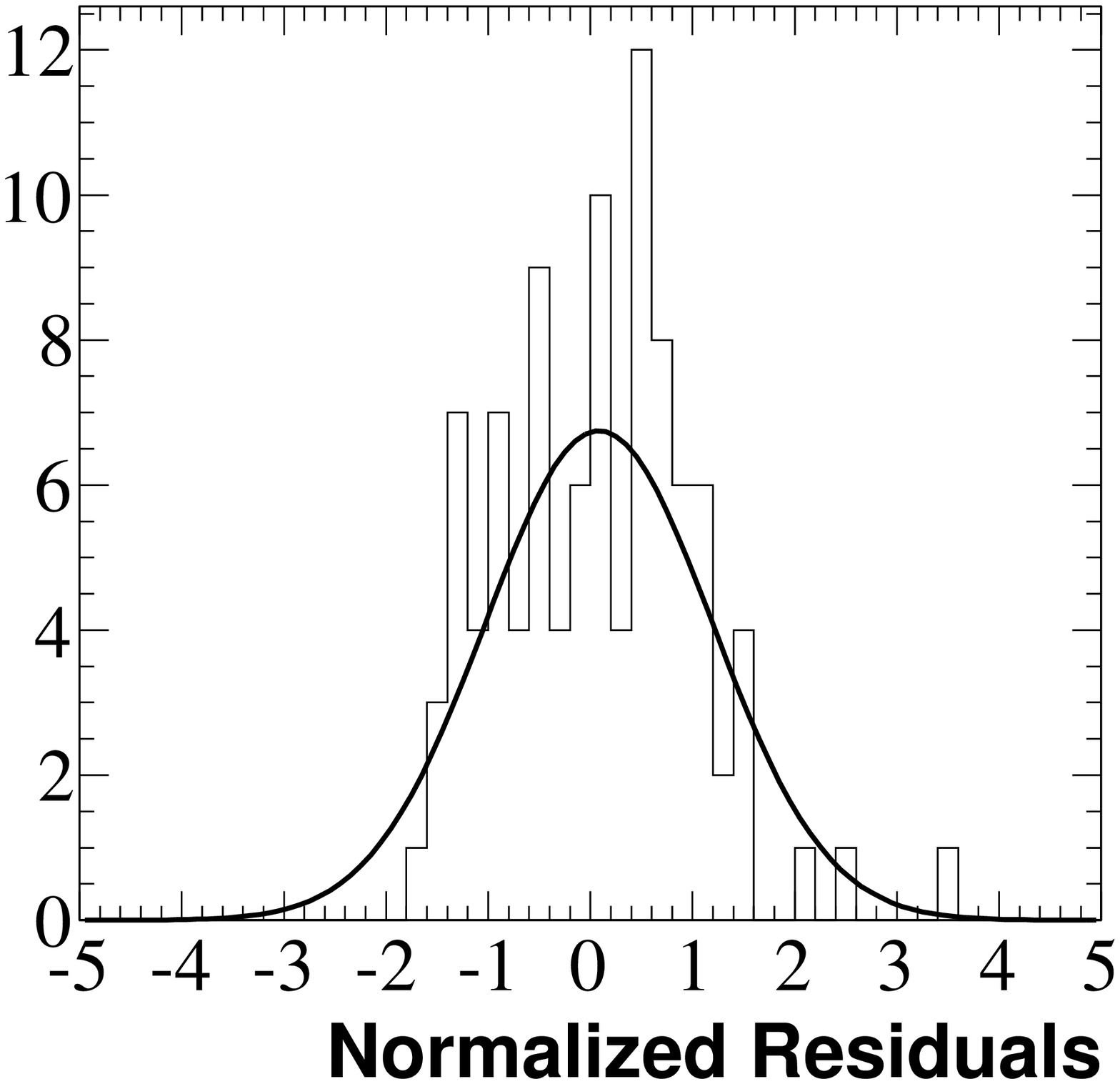}
\vspace{-0.3cm}
\caption{(color online) Normalized residuals of the $D^+$ and $D^-$
  Dalitz plots in equally populated intervals (top) and their
  distribution fitted with a Gaussian function (bottom).}
\label{fig:dpresasym}
\vspace{-0.7cm}
\end{center}
\end{figure}

We pursue a second technique in search of {\CPV}, by measuring
normalized residuals $\Delta$ for the efficiency-corrected and
background-subtracted $D^+$ and $D^-$ Dalitz
plots, where $\Delta$ is defined by
\begin{equation}
  \Delta \equiv {n(D^+) - Rn(D^-) \over \sqrt{\sigma^2(D^+) + R^2\sigma^2(D^-)}},
\end{equation}
with $n(\Dp)$ and $n(\Dm)$ the observed number of \Dp and \Dm mesons in
an interval of the Dalitz plot, where $\sigma(\Dp)$ and $\sigma(\Dm)$
are the corresponding statistical uncertainties.  The results for
$\Delta$ are shown in Fig.~\ref{fig:dpresasym}.  Note that the intervals
for Fig.~\ref{fig:dpresasym} are adjusted so that each interval contains
approximately the same number of events.  We calculate the quantity
$\chi^2/(\nu-1) = (\sum_{i=1}^{\nu}\Delta^2)/(\nu - 1)$, where $\nu$ is
the number of intervals in the Dalitz plot. We fit the distribution of
normalized residuals to a Gaussian function, whose mean and
root-mean-squared (RMS) deviation values we find to be consistent with
zero and one, respectively.  We obtain $\chi^2 = 90.2$ for 100 intervals
with a Gaussian residual mean of 0.08 $\pm$ 0.15, RMS deviation of 1.11
$\pm$ 0.15, and a consistency at the 72$\%$ level that the Dalitz plots
do not exhibit \CP asymmetry.

The Legendre polynomial moments of the cosine of the helicity angle of
the \Dpm decay products reflect the spin and mass of the intermediate
resonant and nonresonant amplitudes, and the interference effects among
them~\cite{Aubert:2007kkpi0}.  A comparison of these moments between the
\Dp and \Dm two-body mass distributions provides a model-independent
method to search for \CP violation in the Dalitz plot, and to study its
mass and spin structure. We define the helicity angle $\theta_H$ for
decays $\Dp\to(r\to K^+K^-)\pi^+$ via resonance $r$ as the angle between
the $K^+$ direction in the $K^+K^-$ rest frame and the prior direction
of the $K^+K^-$ system in the \Dp rest frame. For decays $\Dp\to(r\to
K^-\pi^+)K^+$ via resonance $r$, we define $\theta_H$ as the angle
between the $K^-$ direction in the $K^-\pi^+$ system and the prior
direction of the $K^-\pi^+$ system in the \Dp rest frame.

The Legendre polynomial moment distribution for order $l$ is defined as
the efficiency-corrected and background-subtracted invariant two-body
mass distribution \mkk or \mkpi, weighted by the spherical harmonic
$Y_l^0[\cos(\theta_H)]=\sqrt{2l+1/4\pi}P_l[\cos(\theta_H)]$, where $P_l$
is the Legendre polynomial.  We define the two-body invariant mass
interval weight $W_i^{(l)}\equiv{(\sum_j w_{ij}^{(l)S} - \sum_k
w_{ik}^{(l)B})} / {\langle\epsilon_i\rangle}$, where
$w_{ij}^{(l)}(w_{ik}^{(l)})$ is the value of $Y_l$ for the
$j^{\mathrm{th}}(k^{\mathrm{th}})$ event in the $i^{\mathrm{th}}$ interval
and $\langle\epsilon_i\rangle$ is the average efficiency for the
$i^{\mathrm{th}}$ interval. The superscripts $S$ and $B$ refer to the signal
and background components, respectively.  The uncertainty on $W_i^{(l)}$
is ${\sigma^{(l)}}\equiv\sqrt{{\sum_j (w_{ij}^{(l)S})^2 + \sum_k
(w_{ik}^{(l)B})^2}/{\langle\epsilon_i\rangle^2}}$.  To study differences
between the \Dp and \Dm amplitudes, we calculate the quantities $X_i^l$
for $l$ ranging from zero to seven in a two-body invariant mass
interval, where
\begin{equation}
  X_i^l = \frac{(W_i^{(l)}(\Dp) - RW_i^{(l)}(\Dm))}{\sqrt{{\sigma_i^{(l)}}^2(\Dp) + R^2{\sigma_i^{(l)}}^2(\Dm)}}.
\end{equation}

We calculate the $\chi^2/$ndof over 36 mass intervals in the \kk and
\kpi moments using
\begin{equation}
  \chi^2 = \sum_i\sum_{l_1}\sum_{l_2} X_i^{(l_1)} \rho_i^{l_1l_2} X_i^{(l_2)},
\end{equation}
where $\rho_i^{l_1l_2}$ is the correlation coefficient between $X^{l_1}$ and $X^{l_2}$,
\begin{equation}
\rho_i^{l_1 l_2} \equiv 
\frac{ \langle X_i^{(l_1)} X_i^{(l_2)} \rangle - 
    \langle X_i^{(l_1)}\rangle\langle X_i^{(l_2)}\rangle}
{
    \sqrt{ \langle {X_i^{(l_1)}}^2 \rangle - {\langle X_i^{(l_1)} \rangle}^2} 
    \sqrt{ \langle {X_i^{(l_2)}}^2 \rangle - {\langle X_i^{(l_2)} \rangle}^2} 
},
\end{equation}
and where the number of degrees of freedom (ndof) is given by the
product of the number of mass intervals and the number of moments, minus
one due to the constraint that the overall rates of \Dp and \Dm mesons
be equal.  We find $\chi^2/$ndof to be 1.10 and 1.09 for the \kk and
\kpi moments, respectively (for ndof $= 287$), which corresponds to a
probability of 11\% and 13\%, again respectively, for the null
hypothesis (no \CPV).

\section{\boldmath Model-dependent search for $CP$ violation in the Dalitz plot}
\label{sec:dpamp}
The Dalitz plot amplitude ${\mathcal A}$ can be described by an isobar
model, which is parameterized as a coherent sum of amplitudes for a set
of two-body intermediate states $r$. Each amplitude has a complex
coefficient, \ie, ${\mathcal A}_r[\msqkk,\msqkpi]=\sum_r{\mathcal
M}_re^{i\phi_r}F_r[\msqkk,\msqkpi]$~\cite{CLEO-spin, Asner, BWtext},
where ${\mathcal M}_r$ and $\phi_r$ are real numbers, and the $F_r$ are
dynamical functions describing the intermediate resonances. The complex
coefficient may also be parameterized in Cartesian form, $x_r =
{\mathcal M}_r\cos\phi_r$ and $y_r = {\mathcal M}_r\sin\phi_r$.  We
choose the \Kres~ as the reference amplitude in the \CP-symmetric and
\CP-violating fits to the data, such that ${\mathcal M}_{\Kres} = 1$ and
$\phi_{\Kres} = 0$.

Using events from the sideband regions (defined in
Fig.~\ref{fig:massfit}) of the \Dp mass distribution, we model the \CP
conserving background, which is comprised of the \Kres~and
\phires~resonance contributions and combinatorial background. The
combinatorial background outside the resonant regions has a smooth shape
and is modeled with the non-parametric $k$-nearest-neighbor density
estimator~\cite{knn}. The \Kres~and \phires~regions are composed of
the resonant structure and a linear combinatorial background, which we
parameterize as a function of the two-body mass and the cosine of the
helicity angle. The model consists of a Breit-Wigner (BW) PDF to
describe the resonant line shape, and a first-order polynomial in mass
to describe the combinatorial shape. These are further multiplied by a
sum over low-order Legendre polynomials to model the angular dependence.

\begin{table}[!ht]
\begin{center}
\caption{Resonance mass and width values determined from the isobar model fit to the combined Dalitz-plot distribution.}
\setlength{\extrarowheight}{2pt}
\begin{tabular}{lcc}
\hline
\hline
Resonance & Mass (MeV/c$^2$) & Width (MeV)  \\
\hline
$\bar{K}^{*}(892)^{0}$ & 895.53 $\pm$ 0.17 & 44.90 $\pm$ 0.30 \\
$\phi(1020)$ & 1019.48 $\pm$ 0.01 & 4.37 $\pm$  0.02 \\
$a_{0}(1450)$ & 1441.59 $\pm$ 3.77 & 268.58 $\pm$ 5.28 \\
$\bar{K}^{*}_{0}(1430)^{0}$ & 1431.88 $\pm$ 5.89 & 293.62 $\pm$ 3.83 \\
$\bar{K}^{*}(1680)^{0}$  & 1716.88 $\pm$ 21.03 & 319.28 $\pm$ 109.07  \\
$f_{0}(1370)$ & 1221.59 $\pm$ 2.46 & 281.48 $\pm$ 6.6\\
$\kappa(800)$ & 798.35 $\pm$ 1.79 & 405.25 $\pm$ 5.05\\
\hline
\hline
\end{tabular} 
\label{tab::fit_resparms}
\end{center}
\end{table}

Assuming no {\CPV}, we perform an unbinned maximum-likelihood fit to
determine the relative fractions for the resonances contributing to the
decay: $\bar{K}^*(892)^0$, $\bar{K}^*(1430)^0$, $\phi(1020)$,
$a_{0}(1450)$, $\phi(1680)$, $\bar{K}^*_2(1430)^0$, $\bar{K}^*(1680)^0$,
$\bar{K}^*_1(1410)^0$, $f_2(1270)$, $f_0(1370)$, $f_0(1500)$,
$f_2^{\prime}(1525)$, $\kappa(800)$, $f_0(980)$, $f_0(1710)$, and a
nonresonant (NR) constant amplitude over the entire Dalitz plot. We
minimize the negative log likelihood (NLL) function 
\begin{eqnarray}
-2 \ln \mathcal{L} = \qquad \qquad \qquad \qquad \qquad \qquad \qquad \qquad 
        \nonumber \\ -2 \sum_{i=1}^{N} \ln \bigg[ p(m_{i})
	\frac{\epsilon_{\mathrm{MC}}(x_1,x_2) S(x_1,x_2)} 
	{\iint\epsilon_{\mathrm{MC}}(x_1,x_2) S(x_1,x_2)\mathrm{d}x_1\mathrm{d}x_2} 
	+ \nonumber\\ ( 1 - p(m_{i}) ) 
	\frac{B(x_1,x_2)}{\iint B(x_1,x_2)\mathrm{d}x_1\mathrm{d}x_2}  \bigg ],\qquad
\end{eqnarray}
where $N$ is the number of events.
The reconstructed $D^{+}$ mass-dependent probability $p(m)$ is defined
as $p(m_{i}) = \frac{S(m_{i})}{S(m_{i})+B(m_{i})}$, where $S(m)$ and
$B(m)$ are the signal and background PDFs, whose parameters are determined 
from the mass fit described in
Sec.~\ref{sec:massfit}; $x_1=\msqkk$ and $x_2=\msqkpi$, $S(x_1,x_2)$ is the
Dalitz plot amplitude-squared, $\epsilon_{\mathrm{MC}}$ is the ANN efficiency, 
and $B(x_1,x_2)$ is the \CP-symmetric background PDF.

The mass and width values of several resonances, including the \Kres~and
\phires, are determined in the fit (Table~\ref{tab::fit_resparms}). The
$f_0(980)$ resonance is modeled with an effective BW parameterization:
\begin{equation}
A_{f_0(980)} = \frac{1}{m_0^2 - m^2 - im_0\Gamma_0\rho_{KK}},
\end{equation}
determined in the partial-wave analysis of $D_s^+\to K^+K^-\pi^+$
decays~\cite{delAmoSanchez:2011dskkpi}, where $\rho_{KK} = 2p/m$ with
$p$ the momentum of the \Kp in the $K^+K^-$ rest frame, $m_0 = 0.922
\gevcc$, and $\Gamma_0 = 0.24 \gev$. The remaining resonances (defined
as $r\to AB$) are modeled as relativistic BWs:
\begin{equation}
{\mathrm{RBW}}(M_{AB}) = \frac{F_rF_D}{M_r^2 - M_{AB}^2 - i\Gamma_{AB}M_r},
\end{equation}
where $\Gamma_{AB}$ is a function of the mass $M_{AB}$, the momentum
$p_{AB}$ of either daughter in the $AB$ rest frame, the spin of the
resonance, and the resonance width $\Gamma_R$. The form factors $F_r$
and $F_D$ model the underlying quark structure of the parent particle of
the intermediate resonances.  Our model for the \kpi ${\mathcal S}$-wave
term consists of the $\kappa(800)$, the $\bar{K}_0^*(1430)^0$, and a
nonresonant amplitude. Different parameterizations for this
term~\cite{kpipimipwa, delAmoSanchez:2010d0kshh} do not provide a better
description of data. The resulting fit fractions are summarized in
Table~\ref{tab:ff}. We define a $\chi^2$ value as
\begin{equation}
  \chi^2 = \sum_i^{N_{\mathrm{bins}}} \frac{(N_i - N_{\mathrm{MC}_i})^2}{N_{\mathrm{MC}_i}}
\end{equation}
where $N_{\mathrm{bins}}$ denotes 2209 intervals of variable size. The
$i^{th}$ interval contains $N_i$ events (around 100), and $N_{\mathrm{MC}_i}$
denotes the integral of the Dalitz-plot model within the interval. We find
$\chi^2/$ndof = 1.21 for ndof $= 2165$.  The distribution of the
data in the Dalitz plot, the projections of the data and the model of
the Dalitz plot variables, and the one-dimensional residuals of the data
and the model, are shown in Fig.~\ref{fig:dpfit}.

\begin{table}[!ht]
\begin{center}
\caption{Fit fractions of the resonant and nonresonant amplitudes in the
  isobar model fit to the data. The uncertainties are statistical.
}
\setlength{\extrarowheight}{2pt}
\begin{tabular}{lc}
\hline 
\hline
Resonance & Fraction ($\%$) \\ 
\hline
$\bar{K}^{*}(892)^{0}$ & 21.15 $\pm$ 0.20 \\
$\phi(1020)$ & 28.42 $\pm$ 0.13 \\
$\bar{K}^{*}_{0}(1430)^{0}$ & 25.32 $\pm$ 2.24 \\
NR & 6.38 $\pm$ 1.82 \\
$\kappa(800)$ & 7.08 $\pm$ 0.63 \\
$a_{0}(1450)^{0}$ & 3.84 $\pm$ 0.69 \\
$f_{0}(980)$ & 2.47 $\pm$ 0.30 \\
$f_{0}(1370)$ & 1.17 $\pm$ 0.21 \\
$\phi(1680)$ & 0.82 $\pm$ 0.12 \\
$\bar{K}^{*}_{1}(1410)$ & 0.47 $\pm$ 0.37 \\
$f_{0}(1500)$ & 0.36 $\pm$ 0.08 \\
$a_{2}(1320)$ & 0.16 $\pm$ 0.03 \\
$f_{2}(1270)$ & 0.13 $\pm$ 0.03 \\
$\bar{K}^{*}_{2}(1430)$ & 0.06 $\pm$ 0.02\\
$\bar{K}^{*}(1680)$ & 0.05 $\pm$ 0.16 \\
$f_{0}(1710)$ & 0.04 $\pm$ 0.03\\
$f'_{2}(1525)$ & 0.02 $\pm$ 0.01\\
\hline
Sum & 97.92 $\pm$ 3.09 \\
\hline
\hline  
\end{tabular} 
\label{tab:ff}
\end{center}
\end{table}

\begin{figure*}[!tb]
\begin{center}
\includegraphics[width=0.3\textwidth]{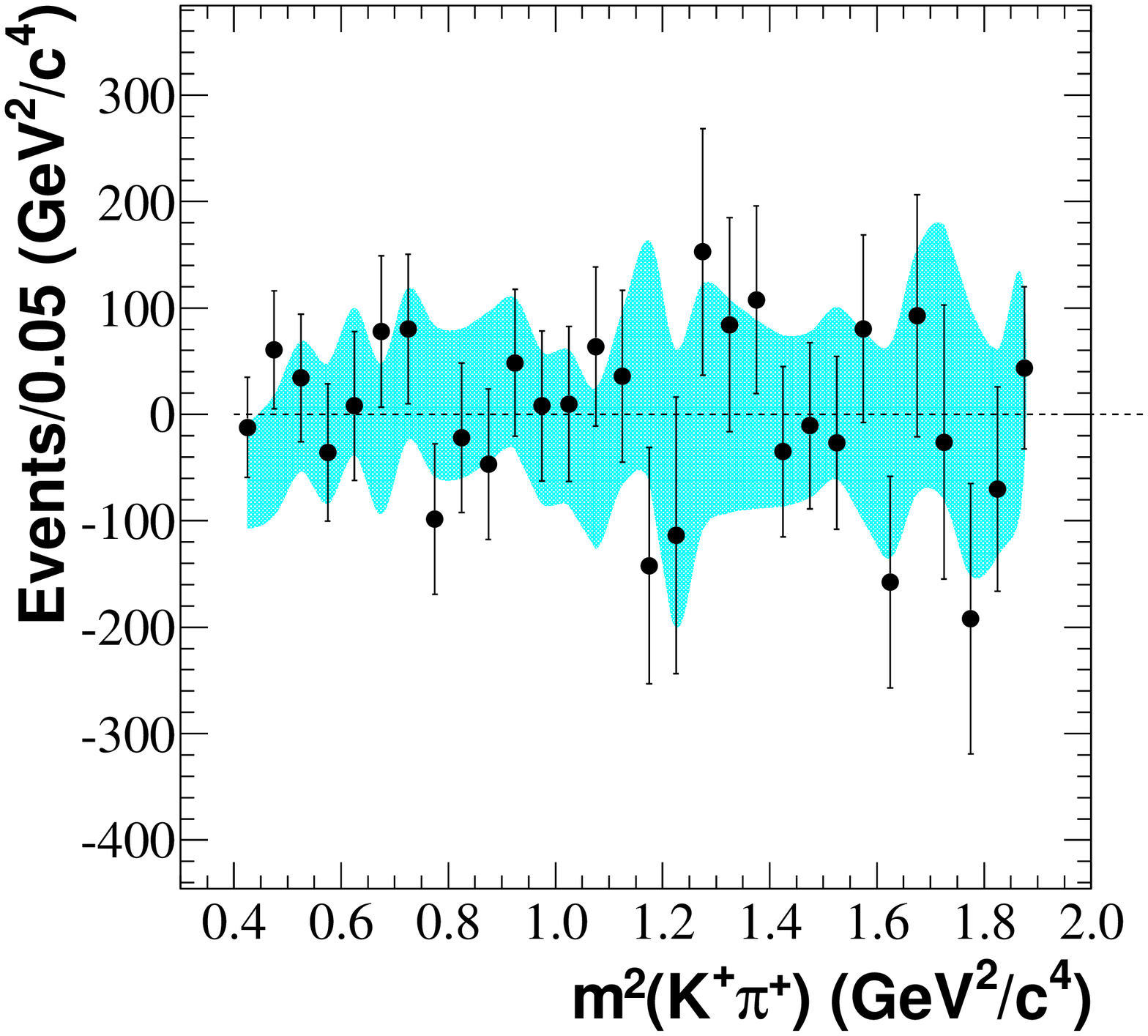}
\includegraphics[width=0.3\textwidth]{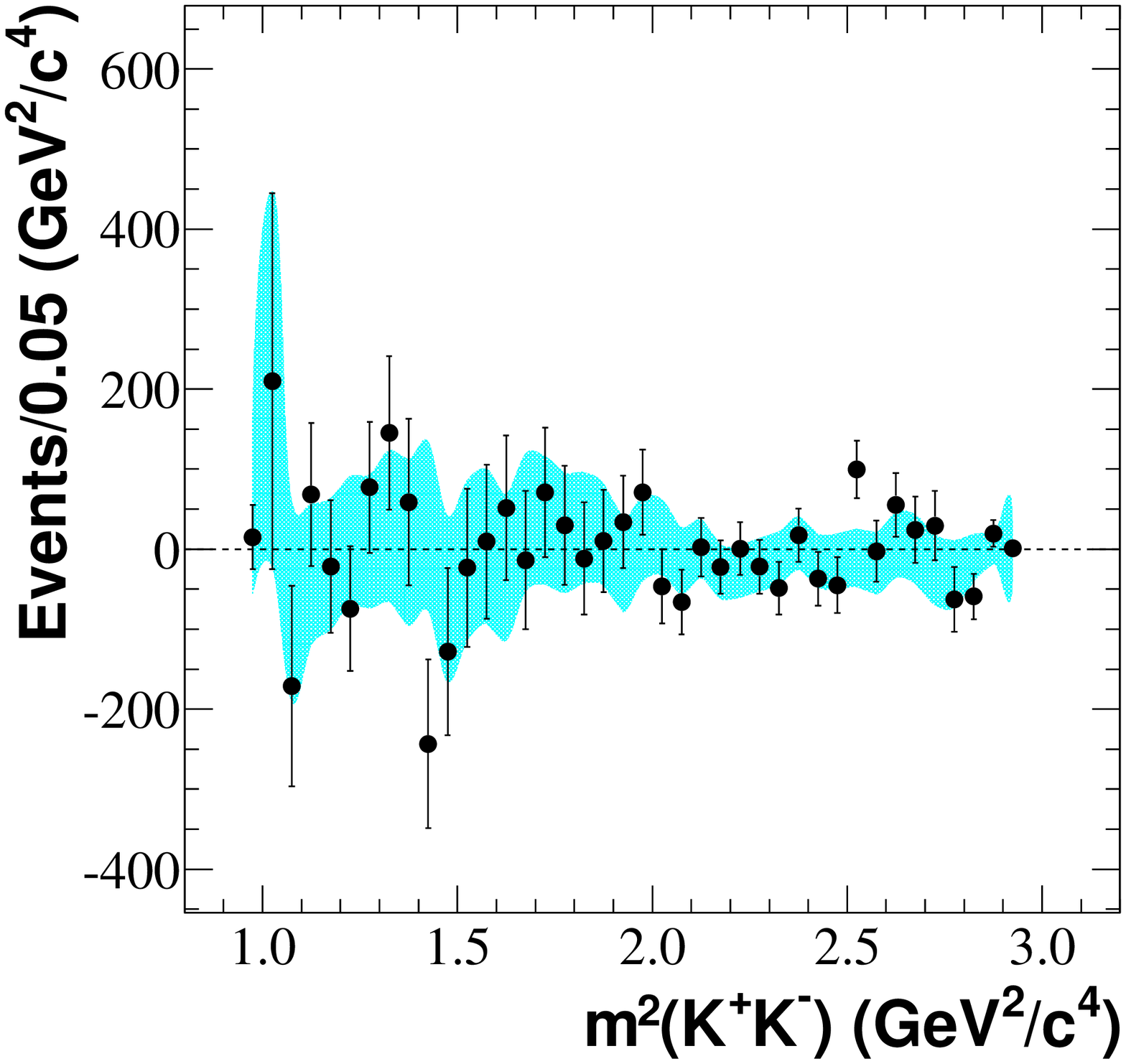}
\includegraphics[width=0.3\textwidth]{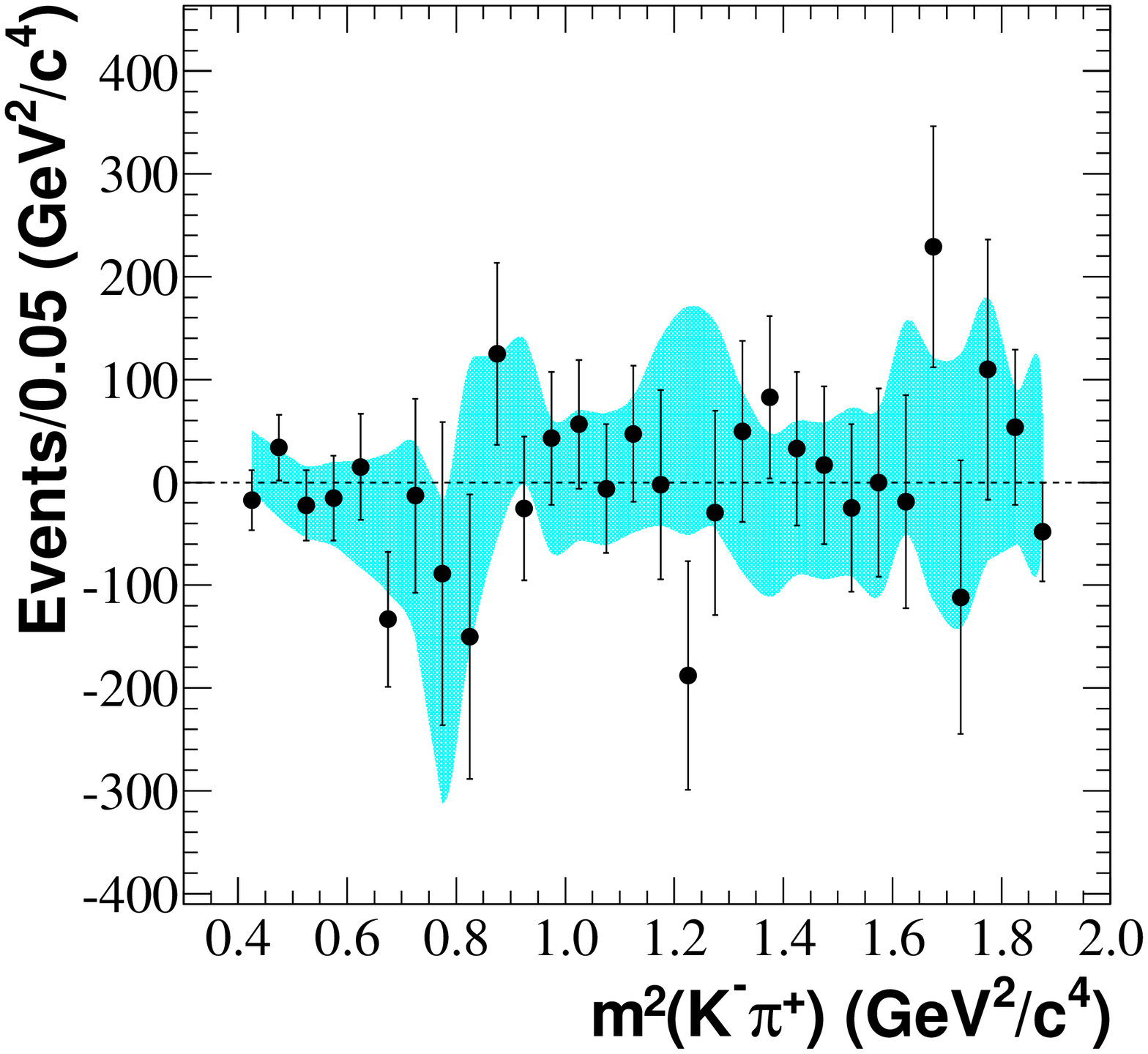}
\vspace{-0.3cm}
\caption{(color online) The difference beween the \Dp and \Dm Dalitz
  plot projections of data (points) and of the fit (cyan band). The
  width of the band represents the $\pm1$ standard deviation statistical
  uncertainty expected for the size of our data sample.}
\label{fig:cpv}
\vspace{-0.7cm}
\end{center}
\end{figure*}

To allow for the possibility of {\CPV} in the decay, resonances with a
fit fraction of at least 1$\%$ (see Table ~\ref{tab:ff}) are permitted
to have different \Dp and \Dm magnitudes and phase angles in the decay
amplitudes (${\mathcal A}$ or $\bar{{\mathcal A}}$). We perform a
simultaneous fit to the \Dp and \Dm data, where we parameterize each
resonance with four parameters: ${\mathcal M_r},\phi_r, r_{CP}$, and
$\Delta\phi_{CP}$. The \CP-violating parameters are $r_{CP} =
\frac{|{\mathcal M}_r|^2 - |\bar{{\mathcal M_r}}|^2}{|{\mathcal M}_r|^2
+ |\bar{{\mathcal M_r}}|^2}$ and $\Delta\phi_{CP} = \phi_r -
\bar{\phi}_r$. In the case of ${\mathcal S}$-wave resonances in the
$K^+K^-$ system, which make only small contributions to the model, we
use instead the Cartesian-form of the \CP\ parameters, $\Delta x$ and
$\Delta y$, to parameterize the amplitudes and asymmetries. This choice
of parameterization removes or eliminates technical problems with the
fit. For these resonances we therefore introduce the parameters
$x_r(D^\pm) = x_r \pm \Delta x_r/2$ and $y_r(D^\pm) = y_r \pm \Delta
y_r/2$. The masses and widths determined in the initial fit (shown in
Table~\ref{tab::fit_resparms}) are fixed, while the remaining parameters
are determined in the fit. In Table~\ref{tab:dpmodel}, we report the \CP
asymmetries, \ie, either the polar-form pair $(r_{CP}, \Delta\phi_{CP})$
or the Cartesian pair $(\Delta x_r, \Delta y_r)$.  Figure~\ref{fig:cpv}
shows the difference between the Dalitz-plot projections of the \Dp and
\Dm decays, for both the data and the fit, where we weight the \Dm
events by the quantity $R$ described in Sec.~\ref{sec:massfit}. It is
evident from the figure that both the charge asymmetry of the data and
fit are consistent with zero and with each other.

\begin{table*}
\centering
\setlength{\extrarowheight}{5pt}
\caption{\CP-violating parameters from the simultaneous Dalitz plot fit. The first
  uncertainties are statistical and the second are systematic.}
\begin{tabular}{lccc}
\hline 
\hline
Resonance & $r_{CP}$ ($\%$) & $\Delta\phi\;(^{\circ})$\\ 
\hline
$\bar{K}^{*}(892)^{0}$          & 0. (FIXED)                                    & 0. (FIXED)                            \\
$\phi(1020)$                    & $0.35^{+0.82}_{-0.82} \pm 0.60$    & $7.43^{+3.55}_{-3.50} \pm 2.35$    \\ 
$\bar{K}^{*}_{0}(1430)^{0}$     & $-9.40^{+5.65}_{-5.36} \pm 4.42$   & $-6.11^{+3.29}_{-3.24} \pm 1.39$   \\
NR                              & $-14.30^{+11.67}_{-12.57} \pm 5.98$ & $-2.56^{+7.01}_{-6.17} \pm 8.91$  \\
$\kappa(800)$                   & $2.00^{+5.09}_{-4.96} \pm 1.85$    & $2.10^{+2.42}_{-2.45} \pm 1.01$    \\ 
$a_{0}(1450)^{0}$               & $5.07^{+6.86}_{-6.54} \pm 9.39$            & $4.00^{+4.04}_{-3.96} \pm 3.83$ \\
\hline
		& $\Delta x$                            & $\Delta y$                    \\              
$f_{0}(980)$                    & $-0.199^{+0.106}_{-0.110} \pm 0.084$ & $-0.231^{+0.100}_{-0.105} \pm 0.079$ \\
$f_{0}(1370)$                   & $0.019^{+0.049}_{-0.048} \pm 0.022$        & $-0.0045^{+0.037}_{-0.039} \pm 0.016$ \\

		    \hline
\hline
\end{tabular} 
\label{tab:dpmodel}
\end{table*}

\section{Systematic Uncertainties}
\label{sec:syst}
We consider the following sources of systematic uncertainty: the
$\mathrm{R}_{\cal{L}}$ selection, corrections applied to the MC, binning
of the data in \coscm, and the Dalitz plot model.

\begin{table}[!ht]
\centering
\caption{Systematic uncertainties for the integrated \CP\ asymmetry.}
\begin{tabular}{lc}
\hline 
\hline
Average $\cos(\theta)$ asymmetry & $\Delta \ACP$[\%] \\ 
\hline
Event selection & 0.07 \\ 
Single forward and backward bin & 0.01 \\
\coscm binning & 0.04 \\
Track asymmetry correction & 0.12 \\
\hline
\hline
\end{tabular} 
\label{tab:syst}
\end{table}

To evaluate the uncertainty due to the $\mathrm{R}_{\cal{L}}$ selection,
we vary the selection such that the yield varies by at least $\pm1$
standard deviation, and assign a systematic uncertainty defined by the
largest variation with respect to the nominal value of the \CP
asymmetry.

The uncertainty due to corrections of the production model in the
simulation (described in Sec.~\ref{sec:corrmc}) is evaluated by randomly
sampling the correction factors from a Gaussian distribution using their
central values and uncertainties as the mean and sigma, respectively.  The
efficiency is then re-evaluated and the fit is re-performed floating the \CP\
parameters while keeping other parameters fixed. This entire procedure
is repeated 50 times. We take the RMS deviation of the 50 fit values of
the \CP\ parameters to obtain the systematic uncertainty estimate. The
uncertainty due to the tracking asymmetry correction is evaluated by
comparing the measurement with two different corrections, namely the
``Tau31'' correction and the correction used in our analysis of $D^+\to
K_S^0\pi^{+}$ decays~\cite{delAmoSanchez:2011kspi}. The average tracking
asymmetry in the latter analysis is $(0.23\pm 0.05)$\%, which is
consistent with the result presented in Sec.~\ref{sec:corrmc} after
accounting for the different momentum spectra. We take the difference
between the \CP\ asymmetry central values using the two different
tracking asymmetry corrections as the systematic uncertainty.

The integrated measurement results from binning the data in \coscm. To
evaluate the effect of the binning in \coscm for the integrated \CP\ 
measurement, we vary the number of intervals and the interval edges, and
measure the \CP asymmetry as the average asymmetry from a single forward
interval and a single backward interval. Systematic uncertainties are 
determined from the difference between the nominal central value and the
value determined from the alternative methods. We report these uncertainties
for the integrated measurement in Table~\ref{tab:syst}. These systematic
uncertainties are combined in quadrature to obtain the final result.

To determine the model-dependent uncertainty on the Dalitz-plot {\CPV}
parameters, we remove resonances with fit fractions less than 1\%, one
resonance at a time, and repeat the fit. We change the standard
value of the radius parameter in the Blatt-Weisskopf form factor
\cite{BWtext} for the intermediate resonance decay vertex from 1.5
GeV$^{-1}$ to 1.0 GeV$^{-1}$. We take the maximum variation as the
model-dependent systematic uncertainty. Systematic uncertainties for the
Dalitz plot fit {\CPV} parameters are listed in Table~\ref{tab:dpmodel}.

Finally, we study possible systematic effects on the binned Dalitz-plot
results presented in Sec.~\ref{sec:moments}. The nominal probability for
the null \CPV\ hypothesis is 72\%\ for 100 intervals, while it is 42\%,
62\%, and 73\%, respectively, for 25, 49 and 144 intervals. In
comparison, changing the $\mathrm{R}_{\cal{L}}$ selection, as described
above, changes the nominal probability to 81\%.

\section{Summary}
\label{sec:sec_sum}
  In summary, we do not find any evidence for \CP violation in the SCS
  decay \Dkkpi. The integrated \CP asymmetry obtained using 
  Eq.~(\ref{eqn:acp}) is $(0.37\pm0.30\pm0.15)\%$. We find that the
  asymmetries in four regions of the Dalitz plot are consistent with
  zero, as listed in Table~\ref{tab::binnedDPAcp_ub}. In addition, the
  \Dp and \Dm Dalitz plots are consistent with no \CP asymmetry with a
  probability of 72$\%$ according to the analysis of the normalized
  residuals for the \Dp and \Dm Dalitz plot divided into 100 equally
  populated intervals. Finally, we find no evidence for \CP asymmetry in
  decays through various intermediate states from a study of the
  two-body mass distributions, as seen in Fig.~\ref{fig:cpv}, and from a
  parameterization of the Dalitz plot for which the \CP asymmetries in
  amplitudes are listed in Table~\ref{tab:dpmodel}.
\section{Acknowledgments}
\input acknowledgements

\end{document}

%% file: authors_jul2012.tex
%
\author{J.~P.~Lees}
\author{V.~Poireau}
\author{V.~Tisserand}
\affiliation{Laboratoire d'Annecy-le-Vieux de Physique des Particules (LAPP), Universit\'e de Savoie, CNRS/IN2P3,  F-74941 Annecy-Le-Vieux, France}
\author{J.~Garra~Tico}
\author{E.~Grauges}
\affiliation{Universitat de Barcelona, Facultat de Fisica, Departament ECM, E-08028 Barcelona, Spain }
\author{A.~Palano$^{ab}$ }
\affiliation{INFN Sezione di Bari$^{a}$; Dipartimento di Fisica, Universit\`a di Bari$^{b}$, I-70126 Bari, Italy }
\author{G.~Eigen}
\author{B.~Stugu}
\affiliation{University of Bergen, Institute of Physics, N-5007 Bergen, Norway }
\author{D.~N.~Brown}
\author{L.~T.~Kerth}
\author{Yu.~G.~Kolomensky}
\author{G.~Lynch}
\affiliation{Lawrence Berkeley National Laboratory and University of California, Berkeley, California 94720, USA }
% ...
\collaboration{The \babar\ Collaboration}
\noaffiliation

%% file: acknowledgements.tex
We are grateful for the 
extraordinary contributions of our \pep2\ colleagues in
achieving the excellent luminosity and machine conditions
that have made this work possible.
The success of this project also relies critically on the 
expertise and dedication of the computing organizations that 
support \babar.
The collaborating institutions wish to thank 
SLAC for its support and the kind hospitality extended to them. 
This work is supported by the
US Department of Energy
and National Science Foundation, the
Natural Sciences and Engineering Research Council (Canada),
the Commissariat \`a l'Energie Atomique and
Institut National de Physique Nucl\'eaire et de Physique des Particules
(France), the
Bundesministerium f\"ur Bildung und Forschung and
Deutsche Forschungsgemeinschaft
(Germany), the
Istituto Nazionale di Fisica Nucleare (Italy),
the Foundation for Fundamental Research on Matter (The Netherlands),
the Research Council of Norway, the
Ministry of Education and Science of the Russian Federation, 
Ministerio de Ciencia e Innovaci\'on (Spain), and the
Science and Technology Facilities Council (United Kingdom).
Individuals have received support from 
the Marie-Curie IEF program (European Union), the A. P. Sloan Foundation (USA) 
and the Binational Science Foundation (USA-Israel).